\documentclass[10pt,journal]{IEEEtran}

\makeatletter
\def\endthebibliography{%
	\def\@noitemerr{\@latex@warning{Empty `thebibliography' environment}}%
	\endlist
}
\makeatother

\IEEEoverridecommandlockouts

\usepackage{cite}
\usepackage{url}
\usepackage{amsmath,amssymb,amsfonts}
\usepackage{algorithmic}
\usepackage{graphicx}
\usepackage{textcomp}
\usepackage{xcolor}
\usepackage{makecell}
\usepackage{multirow}
\usepackage{enumerate}
\usepackage{ulem}
\usepackage{etoolbox}
\makeatletter
\patchcmd{\@makecaption}{\scshape}{}{}{}

\makeatother

\def\BibTeX{{\rm B\kern-.05em{\sc i\kern-.025em b}\kern-.08em
    T\kern-.1667em\lower.7ex\hbox{E}\kern-.125emX}}

\author{He~Xue, Dajiang~Chen,~\IEEEmembership{Member,~IEEE,}
	Ning~Zhang,~\IEEEmembership{Senior Member,~IEEE,} Hong-Ning~Dai,~\IEEEmembership{Senior Member,~IEEE,}
	and~Keping~Yu,~\IEEEmembership{Member,~IEEE}}

\begin{document}

\title{Integration of Blockchain and Edge Computing in Internet of Things: A Survey \\
}

\maketitle

\begin{abstract}

As an important technology to ensure data security, consistency, traceability, \textsl{etc.}, blockchain has been increasingly used in Internet of Things (IoT) applications. The integration of blockchain and edge computing can further improve the resource utilization in terms of network, computing, storage, and security. This paper aims to present a survey on the integration of blockchain and edge computing.
In particular, we first give an overview of blockchain and edge computing. We then present a general architecture of an integration of blockchain and edge computing system.
We next study how to utilize blockchain to benefit edge computing, as well as how to use edge computing to benefit blockchain.
We also discuss the issues brought by the integration of blockchain and edge computing system and solutions from perspectives of resource management, joint optimization,  data management, computation offloading and security mechanism.
Finally, we analyze and summarize the existing challenges posed by the integration of blockchain and edge computing system and the potential solutions in the future.
\end{abstract}


\section{Introduction}

Traditional cloud computing has been used to achieve on-demand resource sharing because of its high flexibility and scalability. With the rapid development of Internet of Things (IoT) in recent years, however, the explosive growth of data has higher real-time requirements for data storage, data processing, and data exchange, which are far beyond the carrying capacity of traditional cloud computing.
Specifically, in many IoT application scenarios, \textsl{e.g.}, Smart Grid and Internet of Vehicles (IoV),
due to large scale data transmission between billions of devices and the data centres, high latency and bandwidth pressure limit the development of traditional cloud computing in IoT.
According to the Telecommunications Industry Association (TIA), the number of connected devices on global network by 2022 will be 29 billion, and roughly 18 billion of them will be related to the IoT \cite{tiaonline}.

As an important complement to cloud computing, a new computing paradigm, named edge computing, has been proposed to expand both computing and storage capabilities from remote clouds to the edge of IoT. In this way, the resources of edge devices can be effectively used to alleviate computing, storage, and bandwidth burdens of traditional cloud computing.
Edge computing can essentially offer distributed and low-latency computing services to smart cities, smart grid, smart healthcare, and other IoT scenarios \cite{sitton2019edge}.
Edge computing can offload computing tasks to edge devices, which are closer to data sources, thereby ensuring privacy preservation and data security \cite{chendajiang2016s2m, chendajiang2018channelprecoding, chendajiang2018physicallayer}. However, due to the limited resources of edge devices, heterogeneity of networks, and highly dynamic environment, many existing data-security techniques cannot be fully utilized in the edge computing architecture \cite{xiao2019edge}.

As an alternative solution to improve the security level and efficiency of edge computing,  blockchain has attracted enough attentions recently \cite{sitton2019review,2020Convergence}.
Blockchain can be regarded as  a decentralized ledger, which utilizes the technologies of peer-to-peer (P2P), cryptography, distributed storage, \textsl{etc.} to achieve the following properties: decentralization, transparency, traceability, security and immutability \cite{wang2019survey}.
In essence, blockchain can enhance the security of edge nodes in edge computing by storing critical data at blockchain \cite{luoguiyang2019software}.
Moreover, blockchain can enable edge computing to implement security mechanisms, such as access control, authentication and privacy preservation, by using well-designed smart contracts \cite{xiong2018mobile,tan2019secure, guo2019trust, nkenyereye2020secure, xiaoyonggang2020edge}.
Furthermore, blockchain can enable edge computing to orchestrate various edge resources through smart contract-based algorithms of resource allocation, task offloading and resource pricing \cite{liu2019blockchain,wang2019permissioned}.
Edge computing, in turn, can support blockchain by providing enough computing resources for the mining tasks.
For instance, when edge devices can provide idle resources as edge servers do, the resources will be allocated in the way of bidding and trading for blockchain.

The integration of blockchain and edge computing (IBEC) is a promising paradigm since both the two technologies can be complementary to construct frameworks to solve problems in several fields.
For instance, in IoV and smart transportation, there exist challenges of insufficient on-board resources of most vehicles to support task processing and data storage, and difficulties in resources allocation brought by the mobility of vehicles \cite{xiao2020daer}.
To this end, the IBEC can serve IoV with collaborative management of computing and communication resources \cite{lin2020blockchain, wang2020consortium, hammoud2020ai, dai2020deep}, data sharing and data management for automatic driving \cite{ li2020vehicle, singh2020blocked, cuilaizhong2020blockchain}, collaborative identity authentication during consensus mechanism \cite{ayaz2020proof, laichenzhe2020security, yanganjia2020delegating, liu2020blockchain}, \textsl{etc.}
In Smart Grid, the IBEC is mainly applied to aspects of pricing and framework designing of energy trading \cite{zhou2019secure, jindal2019survivor, stubs2020blockchain, chen2019framework}, and trading security ensuring \cite{ren2020novel, wang2019blockchain, gai2019permissioned}.
The IBEC can also benefit the other IoT scenarios, such as, Industrial Internet of Things (IIoT) \cite{zhang2019edge, chen2019cooperative, lee2020design, gai2019differential, ren2019identity, xujinlian2019blockchain, queralta2020enhancing, gaoying2020blockchain},  smart healthcare \cite{srivastava4application, tangwenjuan2019fog, akkaoui2020edgemedichain, nguyendinhc2020blockchain, abdellatif2021medge, guo2019access, wangwenming2021blockchain, islam2019bhmus}, edge intelligence and artificial intelligence (AI) \cite{leungvictorcm2021guest, nguyendinhc2021federated, ur2020towards, wang2020ai, zhao2020privacy}, supply chain \cite{husensen2021blockchain, zhang2019knowledge}, smart city \cite{fitwi2019lightweight, xu2019blendmas, hassija2020blockchain}, \textsl{etc.}


Moreover, other technologies (\textsl{e.g.}, cloud computing \cite{armbrust2010view}, fog computing \cite{mahmud2018fog}, Software Defined Network (SDN) \cite{zhang2013cooperative, zhangning2014risk, cheng2013opportunistic}, Network Function Virtualization (NFV) \cite{zhangning2018synergy, renju2016dynamic, zhangning2014dynamic}, AI \cite{chendajiang2020magleak, alelaha2019online, dingyi2020deepedn}), are usually utilized in the architecture of IBEC, either to accomplish specific tasks like video stream processing and model training, or to design adaptable general frameworks for more practical and complicated networks and topologies.
These studies aim to improve the processing performance of various tasks, to increase the resources utilization, and to enhance both network and data security in a more general and realistic environment.

This paper first discusses existing studies including blockchain-enabled edge computing, edge computing-enabled blockchain, and IBEC. Then the challenges and solutions including performance scalability, resource management, security and privacy computing, are summarized and discussed.
Different from other similar studies which view the IBEC as a system \cite{yang2019integrated} or pay attention to the solutions for edge intelligence and blockchain \cite{wangxiaofei2021synergy},
this paper regards blockchain and edge computing as two complementary technologies and focuses on their integrations from the angle of existing issues. The other corresponding technologies are also discussed in this paper only if they contribute to the architecture design and problem solution of IBEC.

 The remainder of this paper is structured as follows.
 In Section \uppercase\expandafter{\romannumeral2}, preliminaries of blockchain, edge computing and IBEC are first reviewed. Then, the issues brought by edge computing and blockchain-enabled solutions are discussed in Section \uppercase\expandafter{\romannumeral3}.
 Section \uppercase\expandafter{\romannumeral4} next discusses the challenges of blockchain and the solutions brought by edge computing.
 After that, the general issues of the IBEC are studied  in Section \uppercase\expandafter{\romannumeral5}.
 Finally, the challenges and solutions for IBEC are presented in Section \uppercase\expandafter{\romannumeral6},  and the paper is concluded in Section \uppercase\expandafter{\romannumeral7}.

\section{An overview of Blockchain and Edge Computing}

This section presents an overview on blockchain and  edge computing technologies in Section \uppercase\expandafter{\romannumeral2}-A and Section \uppercase\expandafter{\romannumeral2}-B, respectively. We then discuss the opportunities brought by the IBEC in Section \uppercase\expandafter{\romannumeral2}-C.

\subsection{Blockchain}

\begin{figure}[ht]
	
	\centering
	\includegraphics[width=9cm]{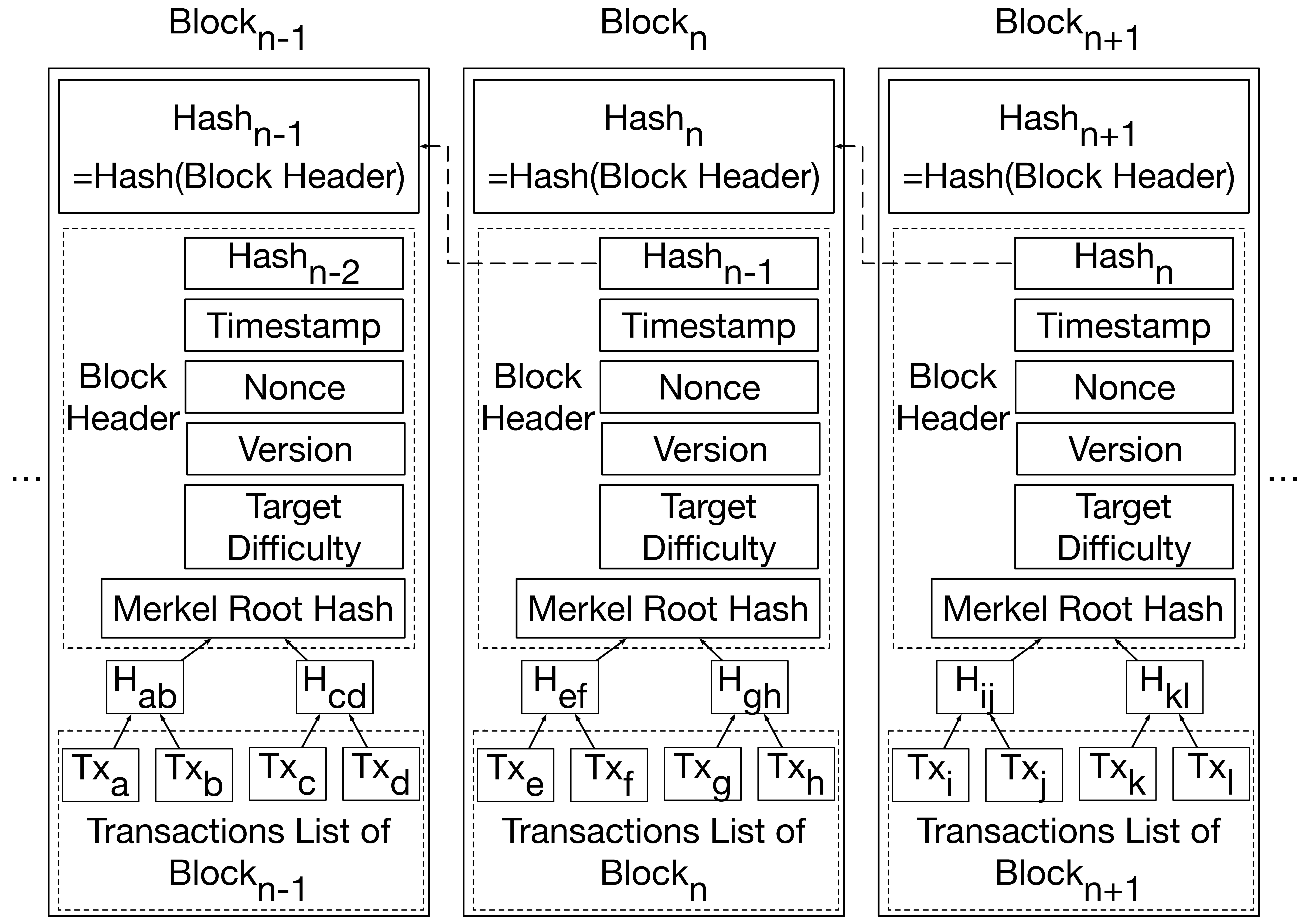}
	\caption{Structure of Blockchain}
	\label{fig:blockchain}
\end{figure}

Blockchain is a kind of chained data structure where data blocks are connected cryptographically and chronologically in sequence for immutability and unforgeability \cite{zhangyuan2019chronos}.
As shown in Fig. \ref{fig:blockchain}, each data block has block header and block body.
The hash value of the previous block, timestamp of the block generating, version, \textsl{etc.} are recorded in the block header. Meanwhile, the transaction list is stored in the block body. Thus, blockchain can be seen as a distributed ledger \cite{liudongxiao2020secureandefficient}.
Blockchain is able to carry valuable digital assets (\textsl{e.g.}, currency, copyrights, contracts, and notarization), and can be used to store or transfer the values of these assets by recording and backtracking because of its immutability and unforgeability. In other words, blockchain can deliver values.
Public blockchains and private blockchains are two main types of blockchain systems \cite{zhangxulin2019blockchainpublicintegrity}.
A public blockchain allows any participant in the network to interact with the records on the chain, while a private blockchain requires only authorized participants to perform corresponding operations.
The characteristics of blockchain are summarized as follows.

%

\begin{itemize}
	
	\item \textsl{Decentralization}: In blockchain, no organization or individual can control global data, and any node that stops working will not affect the overall operation of the system. This decentralized network will greatly improve data security \cite{dai2019blockchain,Yang2019}.
	
	\item \textsl{Tamper-proofing}: In blockchain, the encryption technologies are leveraged to protect data and distributed consensus algorithms are utilized to provide the consistency of data on chain.
Moreover, it requires a large number of nodes (named miners) to participate in verifying transactions and generating blocks,
and modifying any block of the chain requires changing at least half of nodes' subsequent blocks \cite{liudongxiao2019anonymous}; this is nearly impossible for a large scale blockchain network.
	
	\item \textsl{Transparency and Traceable}: The content in the block will be backed up to each node.
All record information is public, and anyone can query the block data through the public interface.
Each transaction is solidified into the block data through chain storage.
All transaction records of all blocks are superimposed in form of hash digests through cryptographic algorithms, which can be traced back to any transaction history \cite{liudongxiao2020transparent,he2020privatefairand,zhangyuan2019blockchain}.	
\end{itemize}


Blockchain is not a brand-new technology, yet a combined innovation that integrates multiple existing technologies as follows.

\begin{itemize}
	
	\item \textsl{Consensus Mechanism}: The goals of consensus mechanism are to build trust among distrusted nodes and obtain rights of generating new blocks. It enables all honest nodes to maintain a consistent view of the blockchain while satisfying properties of consistency and effectiveness. Consistency means that the prefix part of the blockchain kept by all honest nodes is exactly the same.
Effectiveness means the information released by an honest node will eventually be recorded in its own blockchain by all other honest nodes. The commonly-used consensus mechanisms mainly include Proof of Work (PoW), Proof of Stake (PoS), Delegated Proof of Stake (DPoS), Practical Byzantine Fault Tolerance (PBFT), \textsl{etc.}

	\item \textsl{Cryptography}: Cryptography \cite{liu2020blockchainsamrtadvertising} is one of the key technologies used in blockchains.
Many classic modern cryptography algorithms are used in current blockchain applications, including hashing algorithms, symmetric encryption, asymmetric encryption, digital signatures, \textsl{etc.}
	
	\item \textsl{Distributed Storage}: Blockchain is a distributed ledger on P2P network \cite{guan2020towards}.
Each participant will independently store and write block data.
Different from traditional centralized storage, distributed storage has the following two advantages: each node backs up data information to avoid data loss due to single point of failure (SPoF); and data on each node is stored independently to avoid malicious tampering with historical data.
	
	\item \textsl{Smart Contract}:
As a digital protocol to be executed on blockchain, smart contract allows traceable, irreversible and reliable trusted transactions with no third party intervention.
It is essentially a code snippet that includes explicit functions and can be executed automatically when the conditions are met.
Smart contract can be essentially used to enhance transaction security and reduce transaction costs, thereby improving end-to-end cooperation efficiency \cite{lu2019blockchain,miglani2020blockchain}.	
\end{itemize}

\subsection{Edge Computing}

Cloud computing is struggling to deal with the data generated by massive devices in IoT due to high latency and bandwidth cost.
Edge computing provides a nearby service computing paradigm that is close to objects or data sources, which can quickly respond to a variety of services and meet the basic requirements of the industry in real-time business, application intelligence, and security as well as privacy preservation \cite{nijianbing2019toward,khan2019edge}.
The characteristics of edge computing are summarized as follows.

\begin{figure}[t]
	\centering
	\includegraphics[width=9cm]{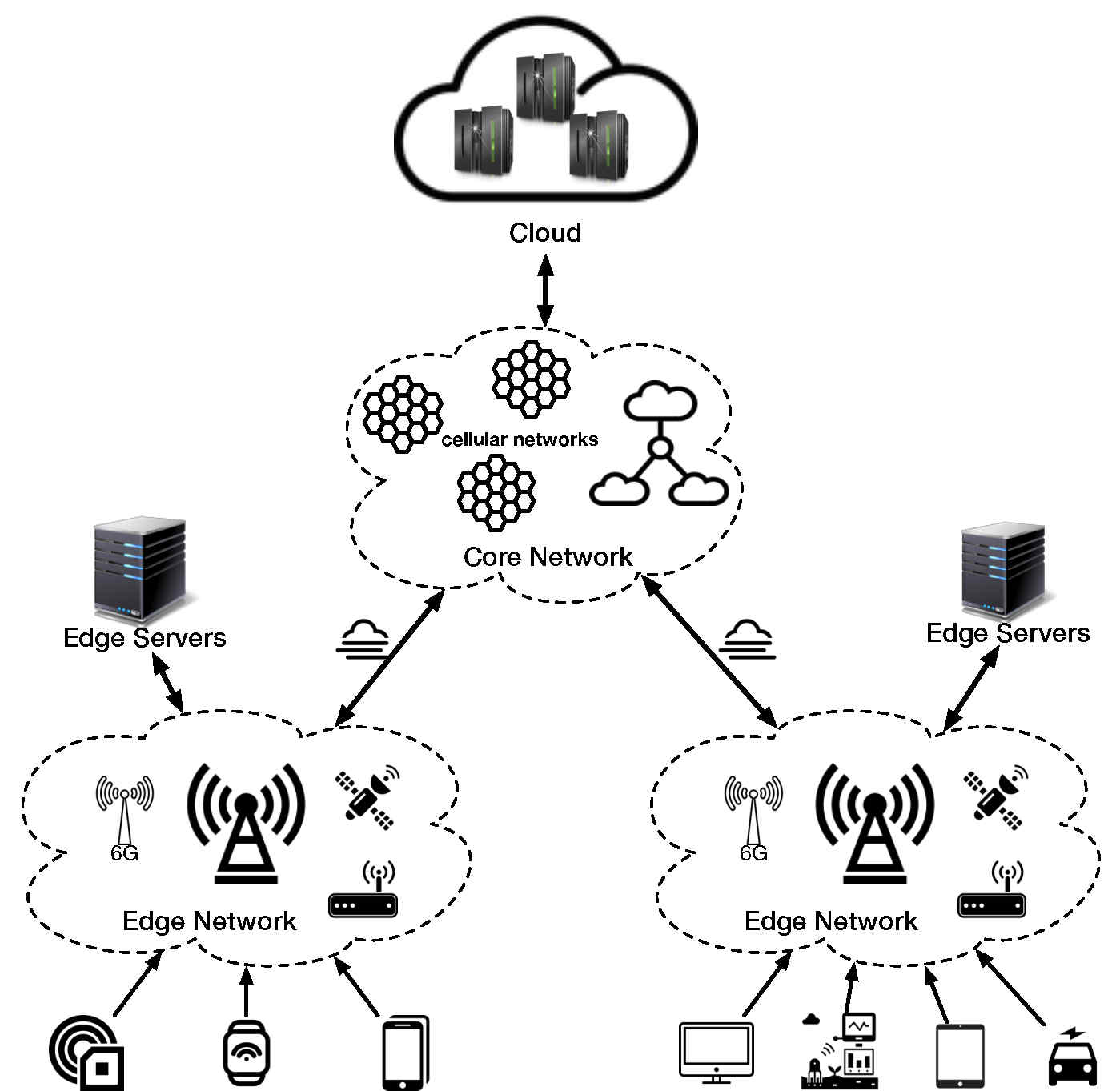}
	\caption{Architecture of Edge Computing}
	\label{fig:edgecomputing}
\end{figure}


\begin{itemize}
	
	\item \textsl{Low Latency}: Edge computing provides computation resources close to the terminals physically and logically.
Therefore, data generation, data processing and data usage all occur within a very close range from the data source, and the latency in receiving and responding to terminal requests is extremely low \cite{shi2016edge,xiaoma2019Cost}.
	
	\item \textsl{Self-organization and High Reliability}: When network interruptions occur, edge severs can achieve local autonomy and self-recovery. With the assistance of edge severs, the central cloud only needs to perform dynamic computation offloading and schedule tasks to specific edge severs.
	
	\item \textsl{Heterogeneous and Scalability}: To fulfill the rising demands for IoT, there are a large number of heterogeneous and scalable edge devices \cite{cheny2020Joint}. Moreover, as a supplement of cloud computing, edge servers in edge computing environment can provide these nearby heterogeneous edge devices more efficient computing, storage and communication services; this means that the resources are moved down from cloud infrastructure to edge side, thus relieving the pressures of all aspects on the cloud layer.
	
	\item \textsl{Low Data Exposure}: Because edge devices can collect and process data locally, it is not necessary to transmit data to remote clouds. Therefore, most information, especially sensitive information, does not need to go through the network, thereby improving security to some extent.
		
\end{itemize}


As shown in Fig. \ref{fig:edgecomputing}, the entire framework of edge computing can be abstracted as a three-level hierarchical architecture as follows \cite{sitton2019review,zhang2018data}.
From the edge to core are the device layer, edge layer and core infrastructure with the increased computing capability and storage capacity. The device layer is composed of large numbers of heterogeneous resource-limited IoT devices with low computing power, \textsl{e.g.}, sensors, Radio Frequency Identification (RFID) tags, cameras, vehicles, roadside units, \textsl{etc.}, which are used to achieve the functions of collecting, transmitting and uploading raw data.
As the servers from the perspectives of computation, edge layer provides storage, computing and network resources to edge devices.
Due to the dynamics of edge server resources and network topology, the computation offloading strategies of tasks and orchestration of corresponding resources are critical to the edge layer.
The core infrastructure generally indicates the cloud layer, whose configurations are similar to cloud computing.
The tasks that cannot be processed by the edge layer need to be completed in the cloud layer.
Moreover, the cloud layer can also dynamically adjust the deployment strategies of the edge layer based on the dynamic allocation of edge resources.

\subsection{IBEC}

\begin{figure*}[ht]
	
	\centering
	\includegraphics[width=\linewidth]{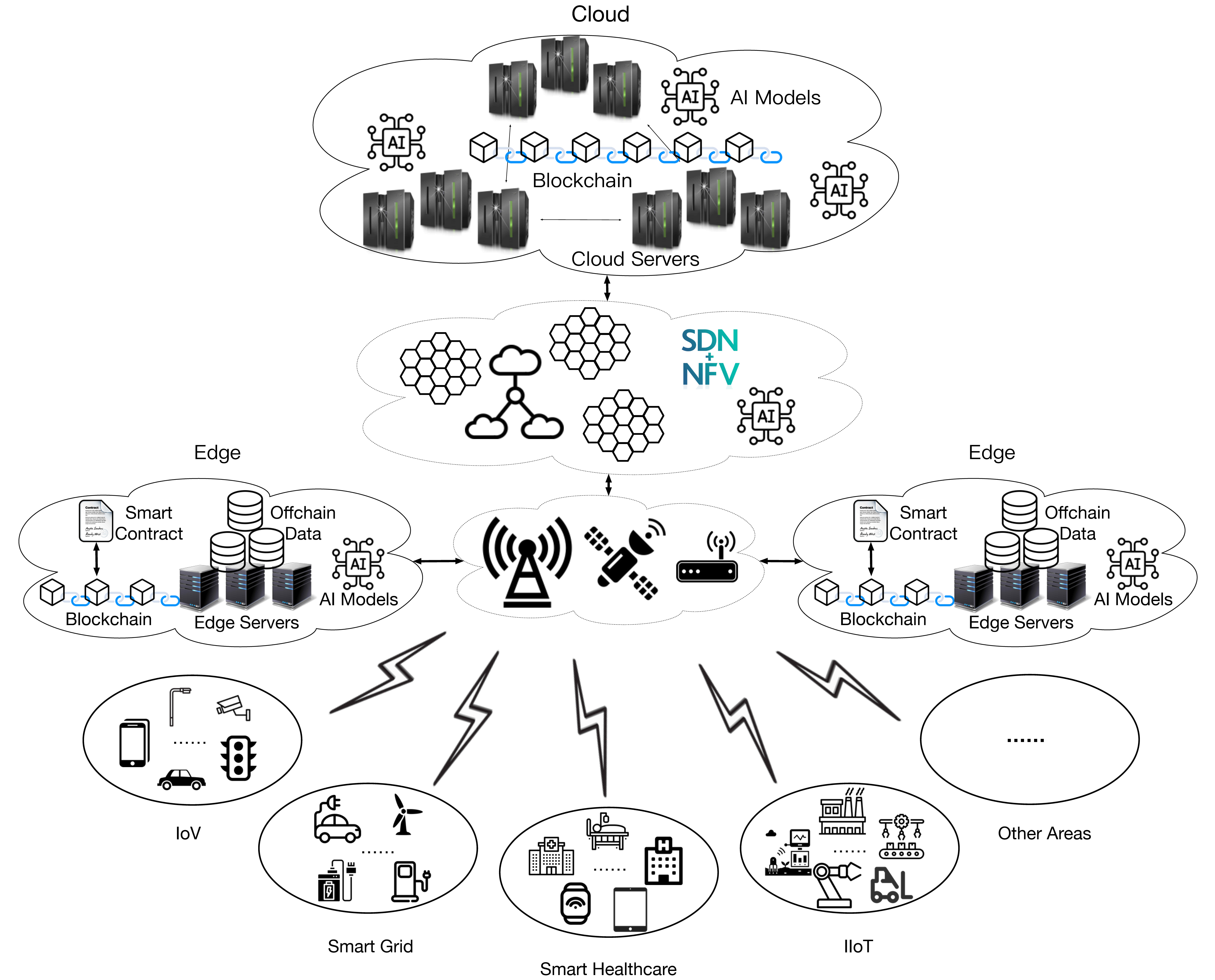}
	\caption{Architecture of IBEC}
	\label{fig:Architecture}
\end{figure*}


The frameworks that integrate blockchain and edge computing have been applied to many IoT scenarios, such as IoV, Smart Grid, and IIoT  \cite{yang2019integrated,fernandez2019towards,luo2020edge,desaimeasuring}.
When blockchain meets edge computing, the privacy preservation, immutability and traceability of blockchain can be leveraged to improve the security of edge computing in IoT. On the other hand, edge computing can be utilized to provide large amount of resources for the high-performance operation of blockchain.
The motivations for combining these two technologies are elaborated as follows.

\begin{itemize}

\item \textsl{Resource Enrichment}: The applications of blockchain in many IoT scenarios (\textsl{e.g.}, data sharing and resource management) require devices deployed on the IoT to  perform consensus algorithms, broadcast or verify transactions, and record historical transactions on blockchain. It is a challenge for IoT devices with limited resources.
Edge computing can provide rich storage, computing and communication resources for edge devices in these scenarios.
    For instance, the deployed edge servers can help the edge devices to process computation-intensive tasks \cite{sitton2019new}.
   Meanwhile, the storage capacity of edge servers can be used to maintain the blockchain for data sharing.
   Edge computing can also provide more powerful network support for edge devices.
	
\item \textsl{Security and Privacy Assurance}: In edge computing, the dynamic topology of networks, and the mobility and heterogeneity of edge devices lead to security challenges \cite{dwang2019PrivStream,2019Blockchainanddeepreinforcement}.
    The blockchain can be used to ensure data consistency, traceability and tamper-proofing, and  protect data privacy \cite{ferrag2018blockchain}. For example, with the effect of the consensus mechanism in blockchain, the participants can jointly maintain data security.
    Moreover, its decentralized architecture based on the P2P network fits the edge computing architecture in IoT.

\item \textsl{Scalability Enhancement}:
The IBEC enables resource-limited edge devices to participate in through certain design mechanisms, although edge servers can be used to assist for processing efficient tasks and maintaining blockchain.
Actually, even resource-constrained edge devices can participate in mining tasks.
For instance, the resource-constrained edge devices can share the collected data to obtain returns with an effective incentive mechanism in blockchain.
Accordingly, the IBEC improves the scalability and flexibility of the entire system.
	
\end{itemize}

The architecture of IBEC is shown in Fig. \ref{fig:Architecture}.
In this architecture,
a variety of resource-limited edge devices connect to the edge network through various access ways, \textsl{e.g.}, 5G, Wi-Fi, and Bluetooth.
The edge servers generally serve as blockchain nodes and miners, because the maintenance of blockchains requires sufficient storage, computing and network resources.
For instance, in energy trading scenario of Smart Grid, edge servers are responsible for executing smart contracts for energy trading requesters.
In data sharing scenario, edge servers train AI models in a distributed manner and finally update the AI models among all nodes through federated learning (FL) paradigm.
On the top of the architecture is the cloud, which is used to deal with some highly complex tasks, which edge nodes cannot complete.
For example, some raw data with high storage overhead (\textsl{e.g.}, X-ray image in healthcare) can be stored in the cloud, while their metadata is stored in blockchain.

\begin{table*}[]
	\caption{Blockchain-based Resource Management for Edge Computing}
	\label{table1}
	\centering
	\renewcommand\arraystretch{1.5}
	\begin{tabular}{l<{\centering}|m{1.7cm}<{\centering}|m{4.0cm}|m{6.3cm}|m{1.5cm}<{\centering}}
		\hline
		Refs. & Applications      & \makecell[c]{Purposes}   & \makecell[c]{Contributions}  & Other Supporting Technologies  \\ \hline
		\cite{zhou2019bcedge} & Resource Management  &  Improving utilization of computational resources and reducing its consumption.    & Proposing a blockchain-based resource management framework named BCEdge and a PoP consensus mechanism. & No            \\ \cline{1-5}
		\cite{qiao2019blockchain} &Resource Management  & Improving execution performance of time sensitive IoT applications. & Proposing a blockchain-enabled D2D-ECN framework and a PoR consensus mechanism.  & No \\ \cline{1-5}
		\cite{pincheira2020rationale} & Resource Management & Avoiding issues introduced by centralized paradigm.  & Implementing a fully decentralized marketplace of fog / edge computing resources based on public blockchain network with functionalities realized by smart contracts. &  No   \\ \cline{1-5}
		\cite{fan2020blockchain} &Resource Management & Effectively managing the license-free spectrum resources in CPSSs. & Proposing a smart contract-based license-free spectrum resource management framework. & No \\ \cline{1-5}
		\cite{seng2020user} & Computation Offloading & Addressing the coordination operations among mobile devices and edge servers in a decentralized and trusted manner.& Designing a blockchain-enabled computation offloading scheme by developing a smart contract-based task-virtual machine algorithm. & Deferred-acceptance algorithm \\  \cline{1-5}
		\cite{luoshunyun2021blockchain} &  Computation Offloading & Achieving high-coverage and audibility of MEC services. & Proposing a private blockchain-based task offloading architecture and three smart contract-based offloading policies. & No \\ \cline{1-5}
		\cite{yuanliang2021coopedge} & Computation Offloading & Enhancing cooperation among edge servers in task offloading. & Implementing a blockchain-based task processing platform with a reputation system and a PoER consensus mechanism. & No \\ \cline{1-5}
		\cite{riveraangelo2020blockchain} &  Computation Offloading & Realizing secure task sharing in MEC. & Proposing a permissioned blockchain-based task sharing framework. & No \\ \hline
		
	\end{tabular}
\end{table*}

\section{Blockchain-Enabled Edge Computing}


In this section, we will review the related studies of blockchain-enabled edge computing. We roughly categorize these studies into blockchain-based resource management for edge computing in Section \uppercase\expandafter{\romannumeral3}-A, blockchain-based security layer in Section \uppercase\expandafter{\romannumeral3}-B, blockchain-based data management for edge computing in Section \uppercase\expandafter{\romannumeral3}-C, and lightweight blockchain for resource-limited edge devices in Section \uppercase\expandafter{\romannumeral3}-D.

\begin{table*}[]
	\caption{Blockchain-based Security Layer for Edge Computing}
	\label{table2}
	\centering
	\renewcommand\arraystretch{1.5}
	\begin{tabular}{l<{\centering}|m{1.7cm}<{\centering}|m{4.0cm}|m{6.3cm}|m{1.5cm}<{\centering}}
		\hline
		Refs. & Applications      & \makecell[c]{Purposes}   & \makecell[c]{Contributions}  & Other Supporting Technologies  \\ \hline
		\cite{zhou2020allstar} &  Resource Management & Realizing a self-organizational, secure and rational Cloud / Edge ecosystem.   & Analysing challenges, introducing ALLSTAR approach including four subsystems and designing business model to motivate the participants.   &  Machine Learning (ML),Cloud Computing  \\ \cline{1-5}
		\cite{tuli2019fogbus} & Resource Management& Facilitating deployment and management of IoT applications and monitoring resource in IoT.  & Proposing a framework of integrating heterogeneous IoT-based systems to fog / cloud named FogBus with detailed modules and services and studying a use case of sleep apnea analysis.    & Cloud Computing, Fog Computing   \\ \cline{1-5}
		\cite{xuxiaolong2019blockchainpowered} & Resource Management  &  Solving the uncertainties caused by resource conflict, service failure and perform degradation in cloud computing. &  Proposing a blockchain and edge computing-based resource provisioning method modelled by a multi-objective optimization problem and solved by NSGA-\uppercase\expandafter{\romannumeral3} for uncertainty-awareness in the workflow.  &  NSGA-\uppercase\expandafter{\romannumeral3} \\ \cline{1-5}
		\cite{sharma2019neural} & Caching Management & Achieving ultra-reliability communications in MEC in terms of availability, connectivity, and survivability.  &  Using a neural network for a blockchain-based intelligent content transport mechanism and smart contracts for unmanned aerial vehicles (UAVs) caching functionalities. & Neural Network, UAV \\ \cline{1-5}
		\cite{wu2020toward} & Trust Mechanism & Guaranteeing trustworthiness in collaborative edge computing process. & Presenting a blockchain-enabled collaborative edge computing called BlockEdge with functionalities of distributed ledger, incentive scheme and reputation mechanism.   &  No  \\ \cline{1-5}
		\cite{jayasinghe2019trustchain} & Trust Mechanism& Optimizing security issues, system delay and resource consumption of traditional blockchain. & Presenting an edge computing-enabled privacy preservation permissioned blockchain named Trustchain for decentralized security concerns.   & No  \\ \cline{1-5}
		\cite{xiao2020reinforcement}  &   Trust Mechanism  &  High processing performance and data security in MEC.  &   Formulating a blockchain-based trust mechanism to resist selfish edge attacks and faked service record attacks, and applying DRL for computational resource allocation decision. &   DRL \\ \cline{1-5}
		\cite{sahasourav2020consortium}  &  Access Control  & Achieving secure communication among various entities in edge-based IoT network.  & Designing a consortium blockchain-based access control scheme to offer mutual authentication between different layers in edge-based IoT environment.  & No  \\ \cline{1-5}
		\cite{bonnah2020decchain}  & Authentication  &  Realizing privacy preservation and data security in edge computing.  &  Proposing a decentralized security scheme named DecChain for user authentication, data storage and data access. &  No  \\ \hline
		
	\end{tabular}
\end{table*}

\subsection{Blockchain-based Resource Management for Edge Computing}
Resource management aims to integrate, allocate and utilize the resources (\textsl{e.g.}, computing, network and storage) in a rational, secure and efficient manner.
An appropriate resource management architecture can largely improve the system performance, thereby enhancing the quality of service (QoS) and quality of experience (QoE) of users.
With respect to edge computing, edge servers offer services for nearby resource-limited edge devices, which involve resource management issues such as resource allocation and computation offloading. Blockchain can enable edge computing to develop efficient resource management architectures, in which smart contracts can be used to implement specific functions, such as resource allocation, reputation establishment and resource trading.

There are several studies on blockchain-based resource management for edge computing. In \cite{zhou2019bcedge}, a resource management scheme, named BCEdge, is proposed, in which three smart contracts are designed to realize the functions of request submission, request response and optimal facility selection of tasks, respectively.
Specifically, a proof-of-performance (PoP) consensus mechanism is presented in BCEdge to support the facility (\textsl{i.e.}, edge device) selection for a task.
Based on the historical token exchange transaction and the contributions that a facility invested to the system, an {optimal facility} can be selected to process tasks in order to relieve the load of edge servers.
Similarly, a blockchain-enabled device-to-device (D2D) edge computing and networks (D2D-ECN) framework is presented in \cite{qiao2019blockchain} with a set of smart contracts for resource trading and task assignment.
For low resource and energy consumption, a proof-of-reputation (PoR) consensus mechanism is proposed in \cite{gai2018proof}.
Moreover, in \cite{pincheira2020rationale}, a fully decentralized resource trading market is designed, in which three smart contracts are designed to trade the fog / edge computing resources in a secure and efficient way.

To address the problem of spectrum contention access and utilize the idle spectrum in Cyber-Physical-Social-Systems (CPSSs), a blockchain and smart contract-based license-free spectrum resource management framework is proposed in \cite{fan2020blockchain} for processing non-real time data in edge / fog-enabled IoT network.
In this framework, the mining-based and auction-based access mechanisms are designed to obtain the access license of spectrum.
Moreover, in order to achieve better performance of the spectrum access,   a key-micro blockchain protocol, named blockchain-KM protocol, is developed to generate two types of blocks.
The key blocks are generated by PoS-after-PoW mechanism and used for recording data of spectrum owners, while micro blocks are generated by lower-PoW mechanism and used for recording transactions.
Two kinds of blocks are connected into a multi-ring structure private blockchain using hash values to ensure that the data on the chain would not be tampered with.

\begin{table*}[]
	\caption{Blockchain-based Data Management for Edge Computing}
	\label{table3}
	\centering
	\renewcommand\arraystretch{1.5}
	\begin{tabular}{l<{\centering}|m{1.8cm}<{\centering}|m{4.0cm}|m{6.2cm}|m{1.5cm}<{\centering}}
		\hline
		Refs. & Applications      & \makecell[c]{Purposes}   & \makecell[c]{Contributions}  & Other Supporting Technologies  \\ \hline
		\cite{zhaofeng2019blockchain} & Data Management & Solving the data security and privacy problems in edge computing environment.  & A blockchain-based data management scheme with the functionalities of data isolation by multi-channel, on-chain data encryption and access control by smart contract.  & No \\ \cline{1-5}
		\cite{li2020preserving} & Data Sharing &  Solving resource limitations problems and guaranteeing data security of data sharing process in intelligent edge.  & Designing an appropriate user-centric blockchain-enabled knowledge sharing framework with a PoP consensus for low energy and low latency. & AI  \\ \cline{1-5}
		\cite{lin2019making} & Data Sharing  & Achieving data trading among heterogeneous edge-AI devices in IoT.  &  Implementing a knowledge trading market with a consortium blockchain as a ledger for secure and efficient transactions management, designing of digital coin and PoT consensus and noncooperative game-based pricing strategy. &  AI   \\ \cline{1-5}
		\cite{cui2019blockchain} & Incentive Mechanism   &  Encouraging content sharing among mobile devices in D2D network.  & Developing two kinds of caching placement schemes considering relationships between allocated computing power and shared data size. &  No   \\ \cline{1-5}
		\cite{zhang2020reliable} & Communication & Guaranteeing the reliability of data transmission in the IoT. & Proposing a data transmission mechanism in edge computing environment based on an optimized blockchain where the edge servers maintain and update the full blockchain, and the general lightweight nodes utilize the data on the chain and the consensus mechanism is redesigned as PoR.   & No \\ \hline
		
	\end{tabular}
\end{table*}

Computation offloading algorithms can be implemented by smart contracts as well.
Aiming at the coordination problem of computation offloading in mobile edge computing (MEC)-enabled ultra-dense wireless networks, a task-VM (\textsl{i.e.}, virtual machine) matching algorithm is introduced in \cite{seng2020user} to match mobile users with suitable edge servers and implemented as a smart contract on blockchain to be performed automatically without trusted third parties.
Considering both coverage of edge services and auditability during task offloading process, a private blockchain-based task offloading architecture is designed in \cite{luoshunyun2021blockchain} for drone-assisted MEC.
In this proposed method, smart contracts with offloading policies are used to select MEC servers for task processing,  in which both the process data and execution results are recorded in blockchain for auditing.
To enhance the task offloading cooperation among edge servers, a blockchain-based decentralized platform in cooperative edge computing, named CoopEdge, is implemented  in \cite{yuanliang2021coopedge}, where a reputation system based on historic task-processing performance of the edge servers is introduced to provide a trustworthy basis for the trustless task offloading environment.
In CoopEdge, the incentive mechanism is employed to motivate more edge servers to participate in processing the peer-offloaded tasks, and a proof-of-edge-reputation (PoER) consensus mechanism is utilized such that the edge server with the highest reputation would be selected as a miner to add new blocks on blockchain.
For the similar purpose, another blockchain-enabled incentive scheme for task sharing among edge servers in MEC environment is proposed in \cite{riveraangelo2020blockchain}.

The above-mentioned studies on blockchain-based resource management for edge computing are summarized in Table \ref{table1}.

\subsection{Blockchain-based Security Layer for Edge Computing}
It is essential to ensure security and traceability in recording status data of resource management in a good resource management architecture.
Blockchain can be used in such architecture to provide a trustworthy environment and ensure data security.
In \cite{zhou2020allstar}, a blockchain-based management architecture is proposed to realize the systematically unified resource management and decentralized ecosystem construction during development and operations (DevOps) process, in which blockchain and smart contracts are deployed to provide trusted services without third party.
Blockchain is also applied in \cite{tuli2019fogbus} to guarantee the data integrity and traceability by recording the hash values of confidential data and providing the services of tracking data.

In order to help generating real-time optimal strategy to resist the uncertainties caused by resource conflict, service failure and performance degradation in the procedure of workflow scheduling in edge computing, a private blockchain is introduced in \cite{xuxiaolong2019blockchainpowered} to record and synchronize the data of real-time strategies and status data of edge nodes, thereby achieving secure and efficient strategies scheduling.
Moreover, a non-dominated sorting genetic algorithm \uppercase\expandafter{\romannumeral3} (NSGA-\uppercase\expandafter{\romannumeral3})-based method is designed to solve the formulated multi-objective optimization problem considering completion time and energy consumption.
In \cite{sharma2019neural}, a blockchain-based approach is designed to ensure connectivity, availability, and survivability, which are fundamental in achieving ultra-reliable and low-latency communication in MEC.

Smart contracts in blockchain can also be used in edge computing to enhance its security  \cite{wu2020toward,jayasinghe2019trustchain,xiao2020reinforcement,sahasourav2020consortium,bonnah2020decchain}.
In \cite{wu2020toward}, a blockchain-powered trusted collaborative edge computing (CEC) service framework, named BlockEdge, is presented
to enhance trustworthiness in CEC.
In BlockEdge, a trust reputation system is constructed based on the deposit balances and verification results on-chain.
In \cite{jayasinghe2019trustchain},  a new blockchain, namely TrustChain, is proposed to establish a trust evaluation mechanism and realize privacy preservation for trust management and data security.
Aiming to prevent attacks of selfish edge server and fake service record, a blockchain-based trust mechanism for offloading task processing in MEC is presented \cite{xiao2020reinforcement}, where a blockchain is deployed to record and manage the trust information, and the consensus protocol combining PoW and PoS is developed. Two versions of edge server central processing unit (CPU) allocation and trust evaluation algorithms are designed by using reinforcement learning (RL) and deep reinforcement learning (DRL). Both the versions are used to reduce the computational consumption of edge servers by optimizing the number of CPUs and to evaluate the service reputation of edge servers using Bayesian inference. Moreover, the latter could further improve the computational performance by introducing modified Boltzmann distribution in achieving the optimal number of CPUs and applying convolutional neural networks (CNNs) to compress the state space.
In \cite{sahasourav2020consortium}, a consortium blockchain-enabled access control scheme in edge computing-based generic IoT environment (CBACS-EIoT) is proposed, which includes the access control phases between devices and gateway nodes as well as those between gateway nodes and edge servers, and the key management phase between edge servers and cloud servers.
In \cite{bonnah2020decchain}, a security scheme  for data storage and access is designed in a fully-decentralized architecture by combining blockchain and edge computing to resist the identity-revealing attack, session hijacking attack, relay attack and man-in-the-middle attack.


The above-mentioned studies on blockchain-based security layer for edge computing are summarized in Table \ref{table2}.

\subsection{Blockchain-based Data Management for Edge Computing}

In edge computing, the operations of massive edge devices will generate a large amount of data, which can be further used for subsequent data analysis and data processing.
The generated data needs to be stored and sometimes shared securely.
The decentralization, immutability, and traceability of blockchain can enable edge computing to achieve secure data management.
In \cite{zhaofeng2019blockchain}, a blockchain-based trusted data management scheme in edge computing, named BlockTDM, is proposed.
BlockTDM supports a range of functions, including mutual authentication for membership management and responsibility tracing, and multi-signature-based PBFT consensus and matrix-based multi-channel data segment for data privacy preservation.
Moreover, user-defined sensitive data encryption, conditional access and decryption query of protected transactions are implemented in  form of smart contracts.

Training by AI models, data can be transferred into valuable knowledge and data trading is equivalent to knowledge sharing to some extent.
Knowledge sharing in edge computing can be enabled by blockchain \cite{li2020preserving,lin2019making,cui2019blockchain}.

In \cite{li2020preserving}, a user-centric blockchain (UCB) framework is presented to deal with the security problems of weak copyright protection, untrusted accounting at edge and inefficient consensus in knowledge sharing.
In UCB, a proof-of-popularity (PoP) consensus mechanism is proposed, which includes user-centric proposal value ranking and electing algorithm and security-aware block generating algorithm.

In \cite{lin2019making}, a blockchain-based knowledge trading market in edge-AI-enabled IoT is presented, in which a consortium blockchain is used to construct the market platform and smart contracts of both knowledge management and knowledge trading are designed to ensure the decentralization, tamper-proofing and confidentiality.
Moreover, a proof-of-trading (PoT) consensus mechanism is proposed by combining PoW and PoS to reduce energy consumption.
Moreover, optimal knowledge pricing strategy is derived as incentive mechanism through Karush-Kuhn-Tucker (KKT) condition that is based on game theory.

Similarly, in order to motivate content sharing in MEC, a blockchain-based solution is proposed \cite{cui2019blockchain}, where base stations serve as edge servers and allocate computing power to run mining tasks to reward cache-enabled sharing mobile devices. By considering linear and nonlinear relationships of computing power and shared data size, optimal caching schemes are developed to maximize total profit and solved respectively by KKT conditions-based solution and difference of convex (DC) programming algorithm.

In \cite{zhang2020reliable}, a blockchain-based data transmission scheme is developed for D2D communication in edge computing,  where a proof-of-reliability (PoR) consensus mechanism is designed such that the users with enough credit and supporters would be allowed to transmit data.
The blockchain is introduced to ensure the tamper-proofing and traceability of data (\textsl{e.g.}, messages and behaviour data) in the IoT.

The above-mentioned studies on blockchain-based data management for edge computing are summarized in Table \ref{table3}.

\begin{table*}[]
	\caption{Lightweight Blockchain for Resource-limited Edge Devices}
	\label{table4}
	\centering
	\renewcommand\arraystretch{1.5}
	\begin{tabular}{l<{\centering}|m{1.8cm}<{\centering}|m{4.0cm}|m{6.2cm}|m{1.5cm}<{\centering}}
		\hline
		Refs. & Applications      & \makecell[c]{Purposes}   & \makecell[c]{Contributions}  & Other Supporting Technologies  \\ \hline
		\cite{pyoung2019blockchain} & Blockchain structure redesign &  Designing appropriate blockchain for edge-based IoT.   &  Proposing a lightweight blockchain called LiTiChain using graphs and two algorithms of inserting a block and deleting / renewing a block.  & No   \\ \cline{1-5}
		\cite{huang2019resource} & Blockchain structure redesign &  Designing appropriate blockchain and consensus mechanism for edge computing.  &  Proposing a redesigned blockchain with an improved PoS mechanism and a recent block storage allocation strategy. & No \\ \cline{1-5}
		\cite{kumar2019proof} & Consensus Mechanism optimization & Achieving an efficient solution for PoW. & Designing an expectation maximization algorithm and polynomial matrix factorization-based approach for PoW.  & Cloud Computing, Fog Computing        \\ \cline{1-5}
		\cite{puthal2019proof} & Consensus Mechanism optimization  & Designing appropriate consensus mechanism for resource-limited edge devices. & Proposing a novel consensus algorithm called PoAh.  & No \\  \hline
		
	\end{tabular}
\end{table*}

\subsection{Lightweight Blockchain for Resource-limited Edge Devices}



When blockchain is used for resource management in edge computing, it requires a large amount of computation resources for consensus algorithms and enough storage resources for storing full blockchain.
Thus, how to support the development of blockchain in edge computing is the priority issue in  blockchain-based resource management.

There are a number of studies on redesigning lightweight blockchains \cite{pyoung2019blockchain,huang2019resource,kumar2019proof,puthal2019proof}.
In \cite{pyoung2019blockchain},  a lightweight blockchain named LiTiChain is proposed to adapt to the edge-enabled IoT system.
In LiTiChain, both transactions and blocks that have a time-limited characteristic are stored in the form of Endtime Ordering Graph (EOG, a data structure of tree in blockchain).
The EOG is based on the order of the end time and an expired block would be deleted to save the storage resources on the edge nodes.
Specific algorithms of $k$-height insertion and deletion / renewal of a block are designed to adjust the structure of  EOG to maintain the order of blocks.
In \cite{huang2019resource}, a novel blockchain is redesigned to suit the edge computing environment,
in which the optimal blocks storage allocation and recent blocks caching strategies are presented to achieve fair storage and quick efficient access of blocks.
Moreover, an improved PoS consensus mechanism is designed for battery-limited edge devices to motivate the nodes to own more tokens and store more data.

An efficient consensus method  is proposed in \cite{kumar2019proof} to solve the mathematical puzzle of PoW consensus by using expectation maximization algorithm and polynomial matrix factorization.
In \cite{puthal2019proof}, a proof-of-authentication (PoAh) consensus algorithm is presented, in which blocks could be packaging by all the nodes and signed by their private keys.
In this proposed algorithm, the selected trusted nodes are responsible for verifying block validation by checking the Media Access Control (MAC) values with the corresponding public key.

The above-mentioned studies on lightweight blockchain for resource-limited edge devices are summarized in Table \ref{table4}.

\section{Edge Computing-Enabled Blockchain}

In this section, we will discuss how to use edge computing to benefit blockchain.

As a new computing paradigm, edge computing can provide sufficient resources for resource-limited edge devices to support blockchain. In other words, edge computing can be regarded as a resource enabler for blockchain.
In \cite{bhattacharya2020mobile},  a MEC-enabled blockchain framework is proposed for IoT, in which mining as a service (MaaS) is designed for IoT devices to purchase computational resources in clouds.
In \cite{xuhaitao2021edge}, a blockchain-based resource allocation scheme is proposed in UAV-enabled edge computing environment, where the resources are allocated between edge computing stations (ECSs) and UAVs, and all data related to the process of resource trading would be recorded in blockchain to achieve tamper-proofing and traceability.
The Multi-leader Multi-follower game is utilized to formulate the interactions between ECSs and UAVs in resource trading process.
The Lagrangian-based solution is designed to solve the profit-maximization problem of ECSs to obtain the optimal resource pricing, and the Bellman dynamic programming-based solutions under open loop situation and feedback situation are employed to address the differential game-based resource demands optimization problem.

\begin{table*}[]
	\caption{Edge Computing-enabled Blockchain}
	\label{table5}
	\centering
	\renewcommand\arraystretch{1.5}
	\begin{tabular}{l<{\centering}|m{1.5cm}<{\centering}|m{4.5cm}|m{6cm}|m{2.5cm}<{\centering}}
		\hline
		Refs. & Applications      & \makecell[c]{Purposes}   & \makecell[c]{Contributions}  & Other Supporting Technologies  \\ \hline
		\cite{bhattacharya2020mobile} & Resource Allocation & Solving PoW puzzles on edge nodes.  &  Reviewing MEC architectures and proposing an edge computing-enabled blockchain framework.    & No  \\ \cline{1-5}
		\cite{xuhaitao2021edge} & Resource Allocation & Optimal resource allocation between UAVs and edge computing servers for blockchain applications.  &  Proposing a resources pricing and trading scheme by using Stackelberg game.  & Game Theory \\ \cline{1-5}
		\cite{jiang2019hierarchical} & Resource Pricing & Maximizing profits of both edge service providers and miners in the process of computation offloading. & Proposing a two-layer computational offloading paradigm and investigating two practical scenarios.     & Game Theory  \\ \cline{1-5}
		\cite{xiong2019cloud} &Resource Pricing & Achieving optimal utilities of miners and profits of resource providers.  & Designing a resource trading scheme by formulating a Multi-leader Multi-follower-based problem and utilizing an ADMM algorithm to solve the problem.  & Game Theory, ADMM algorithm   \\ \cline{1-5}
		\cite{liu2019efficient} & Computation Offloading & Offloading mining tasks to edge nodes. &  Proposing a double auction-based resource allocation scheme for mining tasks and solving it by the step / smooth greedy algorithm-based solution. & No \\ \cline{1-5}
		\cite{guo2020blockchain} & Computation Offloading &  Utilizing idle communication and computational resources on both non-mining-devices and edge clouds for mining tasks.   &  Proposing a resource trading scheme including double auction game-based task offloading to non-mining-devices and Stackelberg game-based task offloading to edge cloud operators.     & Game Theory       \\ \cline{1-5}
		\cite{chang2020incentive} & Resource Pricing   & Achieving optimal profits for both miners and edge service providers.  & Proposing autofit strategies including IR and SR to obtain the maximum profit of the resource pricing problem based on the two-stage Stackelberg game.  & Game Theory        \\ \cline{1-5}
		\cite{rivera2019scalable} & Data Processing & Improving scalability of blockchain and data security in edge computing-based IoT network.  & Implementing an edge computing-based parallel data processing architecture for blockchain mining tasks where blockchain ensures data security.  & No  \\ \hline
		
	\end{tabular}
\end{table*}

Edge computing can be used to derive an optimal resource allocation scheme for blockchain mining tasks.
In \cite{jiang2019hierarchical}, a hierarchical Edge-Cloud computation offloading scheme is presented to obtain optimal resource pricing strategies by using game theory and RL technique.
In \cite{xiong2019cloud},  the Multi-leader Multi-follower game is used to formulate the interactions between miners and cloud / edge providers,
and an Alternating Direction Method of Multipliers (ADMM) algorithm is adopted to cope with the standard Equilibrium Problem with Equilibrium Constraints (EPEC) challenge caused by the high dimensionality of each miner's strategy space, thereby obtaining the maximum profits of providers in consideration of utilities of miners.
Moreover, a combinatorial double auction-based resource allocation scheme and a pricing strategy are designed to achieve budget balance, individual rationality and truthfulness in the auction process \cite{liu2019efficient}.

Non-mining-devices in IoT can also be considered as resource providers.
In \cite{guo2020blockchain}, a collaborative mining network (CMN) is constructed to realize collaborative mining between miners and idle IoT devices or between miners and resource-sufficient edge servers.
The optimization problem between miners and non-mining-devices within a CMN is formulated as a double auction game to calculate the optimal auction price in order to maximize the whole utility of the two sides.
The computation offloading model between edge servers and CMN nodes is formulated into a Stackelberg game to maximize the profits of edge servers, miners and idle IoT devices.
In\cite{chang2020incentive}, two mining schemes: immediate reporting (IR) after successfully computing and strategically reporting (SR) after successfully computing are considered in order to obtain optimal profits of both miners and edge service providers in IR and SR schemes.

In \cite{rivera2019scalable}, an edge computing-based parallel processing framework is designed to improve the scalability of blockchain, where the blockchain is deployed as a decentralized database to synchronize and store the activity data of IoT devices, and
the edge nodes are configured to provide the validator service to perform the parallel processing algorithm.

The above-mentioned studies on edge computing-enabled blockchain are summarized in Table \ref{table5}.

\begin{table*}[]
	\caption{Resource Management for IBEC}
	\label{table6}
	\centering
	\renewcommand\arraystretch{1.5}
	\begin{tabular}{l<{\centering}|m{1.5cm}<{\centering}|m{4.5cm}|m{6.5cm}|m{1.5cm}<{\centering}}
		\hline
		Refs. & Applications      & \makecell[c]{Purposes}   & \makecell[c]{Contributions}  & Other Supporting Technologies  \\ \hline
		\cite{liu2019decentralized} & Resource Management   &  Realizing a secure and efficient video transcoding and delivery approach in distributed environment. & Presenting a smart contract-based MEC network architecture and video transcoding and delivery approach and designing an iteration algorithm as a systematic solution. & Game Theory \\ \cline{1-5}
		\cite{liumengting2020mobile} & Resource Management & Efficient and high-yield video stream transcoding service. & Proposing a blockchain and MEC-enabled transcoding framework with an ADMM algorithm-based optimal resource allocation solution.   & ADMM algorithm  \\ \cline{1-5}
		\cite{sunwen2020joint} & Resource Management & Efficient and appropriate incentives for resource allocation. & Proposing a multi-task cross-server resource allocation framework using IBEC and designing DAMB and BFDA to maximize the system efficiency. & Game Theory \\ \cline{1-5}
		\cite{asheralieva2019distributed} & Resource Management &  Adaptive and efficient communication resource management in blockchain-enabled MEC.  & Proposing a service resource trading framework using blockchain and MEC, and respectively designing a hierarchical RL-based solution and an unsupervised hierarchical deep learning (DL) algorithm for complexity reducing and uncertainties prediction. &  DL, RL, Game Theory   \\ \cline{1-5}
		\cite{asheralieva2019learning} &  Resource Management &  Adaptive and efficient computational resource management in blockchain-enabled MEC.  &  Developing a computational resource trading framework using blockchain and MEC, and designing a hierarchical RL-based interaction decision making approach. &  DL, RL, Game Theory  \\ \cline{1-5}
		\cite{xuqichao2019blockchain} & Caching Management & Secure and efficient edge caching scheme for mobile edge users.  & Proposing a blockchain-based edge caching scheme utilizing smart contracts to record and maintain transactions and designing a pricing mechanism for edge nodes, a gradient descent-based demands seeking algorithm and an optimal caching allocation algorithm for mobile users.  & Gradient Decent algorithm  \\ \cline{1-5}
		\cite{cui2020creat} & Resource Management  &  Improving cache hit rate in edge computing.  &  Designing a blockchain-enabled compressed algorithm of FL named CREAT for training model on decentralized edge nodes and reducing communication load in FL by compressing gradients.  &  FL  \\ \cline{1-5}
		\cite{rathi2020blockchain} & Network Management & Jointly satisfying the trusted and transparent requirements of QoS and optimizing network by resource allocation. & Implementing a Hyperledger Fabric-based multi-domain network management framework with end-to-end network slicing and service-level assurance providing. & 5G, Network Slicing \\ \cline{1-5}
		\cite{rawat2019fusion} & Resource Management & Realizing trusted and transparent relationships, and seamless and dynamic resource exchange in current / future decentralized wireless network.  &  Foreseeing the integration of blockchain, edge computing and SDN for wireless network virtualization and presenting a corresponding architecture with the resistance of double spending attack without the usage of high-speed backhaul links. & SDN  \\ \cline{1-5}
		\cite{muthanna2019secure} & Resource Management & Improving resource utilization and system flexibility and maintaining low latency.  & Proposing a SDN-enabled IoT network architecture to integrate blockchain and edge computing with a task offloading algorithm to offload tasks to OpenFlow switches. & SDN, Fog Computing   \\ \hline

	\end{tabular}
\end{table*}

\section{Existing Issues and the Corresponding Solutions in IBEC}

This section discusses research issues brought by the IBEC.

\subsection{Resource Management for IBEC}

The resource management is the key for integrating blockchain and edge computing to provide diversified services with high quality.
In \cite{liu2019decentralized}, a blockchain-based resource trading system is proposed for video transcoding and is delivered to satisfy various requirements of users in MEC.
In this system, small base stations (SBSs) with MEC servers are regard as miners to generate and verify new blocks. Meanwhile, the video providers work on remote cloud networks and are responsible for video transcoding.
Moreover, mobile users should pay for the  services of transmission and transcoding to edge servers / SBSs and video providers.
The whole resource trading process is self-organized by a series of smart contracts and is modelled into a three-stage Stackelberg game, and a resource allocation iteration algorithm is designed to solve the optimization problem.
To improve the efficiency of video stream transcoding, a MEC and blockchain-enabled transcoding framework is proposed in \cite{liumengting2020mobile}, where the transcoding tasks are divided and offloaded to the nearby D2D nodes or small-cell base stations (SBSs).
For the purpose of achieving maximum average profit of transcoding service, the ADMM-based solution is proposed in form of smart contracts to obtain the optimal block size, offloading scheduling, and computation and spectrum resource allocation.
To maximize the system efficiency of multi-task resource allocation with the constraints of individual rationality, truthfulness and budget balance, two smart contract-based incentive mechanisms are designed in the IBEC system \cite{sunwen2020joint}.
Specifically, a double auction mechanism based on breakeven (DAMB) is proposed to satisfy constraints of all sellers and buyers, and a breakeven-free double auction mechanism (BFDA) is presented to handle more tasks for system efficiency further improvement with less buyers' truthfulness.

In \cite{asheralieva2019distributed},
 a MEC-based blockchain-as-a-service (BaaS) system is proposed to realize efficient resource management and long-term reward in IoT networks,
where UAVs are deployed as aerial base stations (BSs) and work with other terrestrial BSs in the MEC model.
In this system, the interactions between BSs and blockchain nodes are formulated into a stochastic Stackelberg game with multi leaders / BSs and multi followers / blockchain nodes.
The follower's action is regarded as partially observable Markov Decision Process (POMDP) and the leader's random state is modelled as a Markov Decision Process (MDP) according to the optimal responses from followers.
An unsupervised polynomial-time Bayesian deep learning (BDL) algorithm is developed for followers to achieve optimal strategies and an unsupervised polynomial-time deep Q-learning (DQL) algorithm is designed correspondingly for leaders to maximize the expected long-term rewards.
Another related work is \cite{asheralieva2019learning}, in which the resource management of both general computing tasks and mining tasks in IoT applications is studied in order to achieve both the optimal response strategies of miners and the optimal prices of service providers.
%

Besides computation resource, the allocation of caching and network resources is the key for high-quality services in the IBEC.
In \cite{xuqichao2019blockchain}, a blockchain-based trustworthy edge caching scheme is proposed, in which smart contracts are used to record and maintain the transactions and payment information among edge nodes and mobile users.
In this scheme, a pricing mechanism based on historical demands of mobile users for edge nodes,
a gradient descent-based optimal caching demand seeking algorithm for mobile users, and a max-min-based fair caching resource allocation algorithm are designed as a trusted and efficient resource allocation solution to improve the QoE in mobile cyber-physical system (MCPS).

With the purpose of raising caching hit rate in edge computing environment, a blockchain-assisted compression algorithm of FL for content caching named CREAT is proposed in \cite{cui2020creat}, which includes the FL-based proactive content caching algorithm (FPCC) and compression algorithm.
The FPCC algorithm is designed for edge nodes to train the models using local data and predict caching contents according to the popularity and the trained models.
The compression algorithm is applied in FPCC to improve communication efficiency of FL by compressing the uploaded gradients.
Moreover, four smart contracts are designed to interact with the deployed blockchain to ensure data security.
Based on Hyperledger Fabric, a multi-domain edge network resource management framework is implemented in \cite{rathi2020blockchain}, where service-level agreements (SLAs) are written in smart contracts and stored on blockchain.
The data of escrow accounts, transaction number and domain edge orchestrator (DEO) ID is received by multi-domain edge orchestrator (MDEO) and recorded on blockchain, which not only supports network slicing resources provisioning algorithm to perform in a trustworthy and transparent manner, but also satisfies the SLAs, thus improving the QoS.

In \cite{rawat2019fusion}, an architecture of fusion of blockchain, edge computing and SDN is presented to make a foreseeing on wireless network virtualization, where blockchain and edge computing are designed  to resist the double-spending attack without the usage of the high-speed backhaul links.
The architecture can be used to realize trusted yet transparent relationships, and resource exchange in current / future decentralized wireless network seamlessly and dynamically. In \cite{muthanna2019secure}, edge / fog computing and SDN are introduced into the IoT network to develop a high flexible framework of IoT system, and blockchain is employed to ensure the reliability, availability and scalability of the overall system.

The above-mentioned studies on resource management for IBEC are summarized in Table \ref{table6}.

\begin{table*}[]
	\caption{Joint Optimization for IBEC}
	\label{table7}
	\centering
	\renewcommand\arraystretch{1.5}
	\begin{tabular}{l<{\centering}|m{1.7cm}<{\centering}|m{4.3cm}|m{6.8cm}|m{1.5cm}<{\centering}}
		\hline
		Refs. & Applications      & \makecell[c]{Purposes}   & \makecell[c]{Contributions}  & Other Supporting Technologies  \\ \hline
		\cite{feng2020joint} &  Resource Management  &  Achieving optimal performance balancing between blockchain system and MEC system.   &   Formulating an optimization framework with IBEC by jointly considering user association, data rate allocation, block producer scheduling and CPU-cycle frequencies and designing corresponding algorithms for minimization of both energy consumption and delay / time.   &  No  \\ \cline{1-5}
		\cite{guo2019adaptive} &  Resource Management  & Intelligent and secure resource management in future wireless network.  &   Proposing an integrated framework of blockchain and edge computing with PBFT and PoS-based consensus mechanisms and DRL-based solution for the optimization problem of system performance by jointly considering spectrum allocation, block size and the number of consecutive blocks.    & Double-dueling DQN  \\ \cline{1-5}
		\cite{qiu2020networking} &  Resource Management  &  Efficient integration of ML into edge computing-based IoT. &   A collective learning approach by integrating DQL and blockchain and an improved consensus mechanism of PoL are jointly employed on solving resource allocation problems in networking integrated cloud-edge-end architecture.  &  DQL  \\ \cline{1-5}
		\cite{li2020uav} &  Communication   &  Maintaining data transmission in a secure and reliable manner in damaged M2M communication networks.   &  Proposing a blockchain and edge computing-enabled M2M communication framework using UAVs for communication connectivity and designing a dueling DQN-based optimizing algorithm to improve data computation capacity and throughput of the blockchain.  &  Dueling DQN   \\ \cline{1-5}
		\cite{li2020resource} &  Resource Management & Satisfying requirements of processing power, security and performance for delay-tolerant data in M2M communication networks.  &  Introducing a blockchain and edge computing-based system performance joint optimization framework and selecting dueling DQN as the solution of the formulated optimization problem by considering caching, computation and blockchain security.   &  Dueling DQN \\ \cline{1-5}
		\cite{zhang2020deep} & Incentive Mechanism  &  Motivating users to participate in idle resource sharing.  &  Constructing a smart contract-based D2D cache and delivery market with pPBFT as the consensus approach and blockchain as the decentralized ledger and designing a DRL-based solution for caching placement and smart contract performing nodes selection. &  DRL  \\ \cline{1-5}
		\cite{feng2019cooperative} & Computation Offloading &  Secure and efficient computation offloading and resource allocation schemes in MEC.  &  Proposing a blockchain-enabled resource managing joint optimizing framework with an asynchronous advantage actor-critic RL algorithm as the solution for maximizing computation rate of MEC system and transaction throughput of blockchain system.  & DRL, RL \\ \hline
		
	\end{tabular}
\end{table*}

\subsection{Joint Optimization for IBEC}

If the IBEC is regarded as an overall system and they are regarded as two subsystems, there exist joint optimizations for both subsystems.
In \cite{feng2020joint}, an allocation framework of radio spectrum and computational resources is proposed based on blockchain and MEC.
The constraints of user association, data rate allocation, CPU-cycle frequencies of both offloading tasks and blockchain system, and block producer scheduling are jointly considered to obtain an optimal balance between the energy consumption of MEC and the delay / time to finality (DTF) of blockchain.
The above optimization problem is formulated into a mixed-integer nonlinear programming (MINLP) problem.
In \cite{guo2019adaptive}, a resource allocation approach for wireless networks in blockchain-enabled mobile edge computing (B-MEC) framework is proposed, in which the consensus mechanism basing on PBFT is deployed.
In this approach, by considering spectrum allocation, block size and the number of consecutive producing blocks per producer, a joint optimization problem is formulated as a MDP to simultaneously obtain the optimal performance of both MEC and blockchain.
The performance measurement indexes include average execution delay of MEC and the throughput and time to finality / confirmation latency of blockchain.
A DRL-based algorithm is designed by using double-dueling deep Q-network (DQN) to solve the above-mentioned problem.

To solve the joint optimization problem of integrated networking and computing resource allocation, a blockchain-enabled DQL approach, named collective Q-learning (CQL), is proposed in \cite{qiu2020networking}.
In this approach, parts of the learning models are trained by the decentralized IoT nodes locally, and blockchain is used to record and share learning results in a secure, reliable and auditable manner.
Instead of PoW, the designed proof-of-learning (PoL) consensus mechanism allows  the nodes (who train the Deep Neural Network, namely DNN, with minimum reduced percentage of learning loss function) to generate the new blocks and even to trade these if they want.
Utilizing the similar technologies, in \cite{li2020uav},  a resource allocation scheme is proposed for data transmission process in UAV-assisted machine-to-machine (M2M) communications to maximize the data computation capacity and throughput of blockchain.
In this scheme, the blockchain is utilized to ensure data security and privacy, and the MEC is used to improve computing power.
The optimization problem of resource allocation is formulated and solved by dueling DQN to achieve the optimal strategy including data transmission request, computing node selection, block size decision and block interval decision.

In \cite{li2020resource}, a blockchain and edge computing-enabled resource joint optimization framework is presented for delay-tolerant data in M2M communications, in which the data is transmitted periodically and can tolerate a relatively long latency.
In order to solve the time-varying problem,  the DQN-based approach is designed and the optimal selections and decisions on caching servers, computing servers and blockchain system are obtained to further satisfy the requirements of system rewards.
Other related studies include cooperative computation offloading and resource allocation \cite{feng2019cooperative}, and the D2D and MEC caching \cite{zhang2020deep}, in which DRL is used to develop solutions for the corresponding joint optimization problems.
Moreover, in \cite{zhang2020deep}, the partial Practical Byzantine Fault Tolerance (pPBFT) consensus mechanism is proposed for smart contract execution nodes to decide whether or not to execute and push one transaction, and the integrated DPoS and aBFT mechanism is proposed to reach the block producing consensus.

The above-mentioned studies on joint optimization for IBEC are summarized in Table \ref{table7}.

\subsection{Data Management Framework for IBEC}

Different from resource management, 
data management focuses on the secure and efficient access, sharing, and tracing the data generated from edge computing-enabled IoT environment.
Blockchain can provide a decentralized and third-party-free environment in edge computing based on the characteristics of the underlying P2P network.
Accordingly, the decentralization and self-organization of a data management system can be guaranteed by using blockchain.
In \cite{xu2020edgence}, a blockchain and AI-enabled edge computing platform, named Edgence, is proposed for intelligent management of massive decentralized applications (dApps) in IoT.
In Edgence, a three-tier verification including smart contract verification is designed to provide different ways for various demands of IoT-based dApps.
Moreover, decentralized crowdsourcing and decentralized AI training are integrated to realize intelligent management without data transmission to cloud to reduce latency and protect data privacy.
To manage data generated by IoT devices, a three-layer data management architecture (\textsl{i.e.},  the device layer, fog layer and cloud layer) is proposed  in \cite{el2019architecture}, which combines blockchain, edge computing, SDN and NFV.
The smart contract-based access management algorithm and orchestration management algorithm are designed to automatically handle requests from IoT devices without any third party, and the combination of NFV and SDN are used to optimize the performance of resource allocation.

\begin{table*}[]
	\caption{Data Management Framework for IBEC}
	\label{table8}
	\centering
	\renewcommand\arraystretch{1.5}
	\begin{tabular}{l<{\centering}|m{1.5cm}<{\centering}|m{4.5cm}|m{6.8cm}|m{1.5cm}<{\centering}}
		\hline
		Refs. & Applications      & \makecell[c]{Purposes}   & \makecell[c]{Contributions}  & Other Supporting Technologies  \\ \hline
		\cite{xu2020edgence} & Data and Resource Management & Efficient, intelligent and cost-cutting management of heterogeneous decentralized applications in IoT jointly using edge clouds and blockchain.  & Proposing a dApps management platform with three-tier validation for better demand satisfaction and AI training based on blockchain and edge computing.   & DL   \\ \cline{1-5}
		\cite{el2019architecture} & Data and Resource Management  & Convenient and efficient access, managing and processing mass IoT data and secure resource management. & Proposing a hybrid architecture using blockchain, edge computing, cloud computing, SDN and NFV together and designing the corresponding algorithms.   & SDN, NFV      \\ \cline{1-5}
		\cite{nyamtiga2019blockchain} & Data Management & Achieving a secure and practical decentralized data storage solution in edge computing. &  Proposing a solution framework for anonymity, integrity and adaptability of data storage by reviewing the challenges and requirements of deployment of blockchain and edge computing in IoT systems and implementing a system prototype utilizing Ethereum's JavaScript API. & No \\ \cline{1-5}
		\cite{ren2019secure} &Data Management& Solving security of data stored in edge computing environment.  & Proposing a storage architecture with global blockchain deployed on the cloud layer and local lightweight blockchain running on IoT devices for storage of data hashes.  & No  \\ \cline{1-5}
		\cite{nawaz2020edge} & Data Management  & Secure and efficient data trading environment in IoT.  &  Presenting a data trading platform using Ethereum with consensus of proof-of-concept (PoC).  &  No  \\ \cline{1-5}
		\cite{nawaz2019edge} &  Data Management    & Security and privacy-assured data management framework for privacy-critical systems.      &  Implementing a private Ethereum blockchain-based edge computing architecture using AI for data analysis on edge nodes locally and the learned data is shared by smart contracts. &  Ethereum, AI      \\ \cline{1-5}
		\cite{wangxiaoding2020secure} &  Data Management  & Secure and efficient data sharing in edge-enabled IoT. &  Developing a blockchain-based data aggregation strategy and designing a DRL-based self-adaptive double bootstrapped deep deterministic policy gradient solution. &    DRL  \\ \hline
		
	\end{tabular}
\end{table*}

Since a large amount of  sensitive data is required to store in the IBEC system securely, secure storage is a key issue in data management.
In \cite{nyamtiga2019blockchain}, edge servers integrated with blockchain are utilized to provide large storage securely, in which the issues of integrity, adaptability and anonymity are addressed separately by Data Integrity Service (DIS), off-chain state channels (provided by Lightning networks or Raiden networks) and a hybrid cryptographic mechanism with linkable ring signature and zerocash techniques.
In \cite{ren2019secure}, a  secure data storage scheme in edge computing is proposed by using blockchain and regeneration code to cope with the challenges brought by resource-limitations of sensors and data diversities.
Specifically, based on the general three-tier edge computing architecture, a global blockchain is deployed at the cloud layer to store all the data, and regenerative code technology is integrated to provide data redundancy so as to improve data reliability.
Moreover,  a local lightweight blockchain is maintained by edge servers at the edge layer and is deployed on the IoT layer.
The data transmission mechanisms between IoT devices and the local blockchain are designed to periodically synchronize data in order to ensure data integrity.

In the IBEC, blockchain can be used for data trading or sharing in a decentralized manner. This is mainly because the incentive mechanism in blockchain can motivate data producers to share high-quality data \cite{nawaz2019edge,nawaz2020edge,wangxiaoding2020secure}.
For instance, a four-layer data trading architecture, named EdgeBoT, is introduced in \cite{nawaz2020edge} for the IoT system by combining edge computing and Ethereum blockchain.
In this architecture, edge gateways are designed to run AI algorithms, blockchain and smart contracts.
Due to the built-in security mechanisms of blockchain, the architecture can be used to realize data processing and decision making locally, thereby reducing the bandwidth consumption of data uploading.
In \cite{wangxiaoding2020secure}, a blockchain-enabled secure data aggregation strategy (BSDA) is proposed for edge computing-based IoT.
In this method, a new block generation algorithm is designed to improve the throughput and reduce transaction latency.
The labels of security level and task completion requirements are put into the block header to restrain the task receivers, \textsl{e.g.}, mobile data collectors (MDCs).
Moreover, integrated with MDCs partition, sensitive task decomposition is adopted to preserve the privacy and to resist the collusion attack simultaneously.
Besides, with the higher MDC configurations of security level and task completion conditions than those recorded in block header, a DRL-based self-adaptive double bootstrapped deep deterministic policy gradient (IDDPG) method is developed to achieve data aggregation with energy-efficiency.

The above-mentioned studies on data management framework for the IBEC are summarized in Table \ref{table8}.

\subsection{Computation Offloading for IBEC}
Computation offloading refers to transferring the entire or part of a task to nearby edge servers to execute for resource-constrained edge devices. During the transferring process, data about the tasks is vulnerable to be attacked in the sophisticated environment. In this regard, blockchain can be used as a ledger to ensure data consistency and integrity by recording the status data of resource allocation and task offloading on-chain.

In \cite{cui2019decentralized}, a blockchain-enabled edge computing system, named DeTEC, is designed by utilizing the idle computing resources for task offloading in a decentralized and trusted manner.
Taking capacity and fairness of the edge servers into consideration, the optimal task allocation problem is proposed in order to minimize the total latency, and is solved by the heuristic allocation (Heu-Alloc) algorithm.
Specifically, when the task requests from IoT devices are collected, the Heu-Alloc algorithm is run to generate a scheduling scheme, which is sent to the smart domain name server (DNS).
The appropriate edge server is selected to process the corresponding tasks and return the results to IoT devices.
Moreover, two verification schemes are designed to improve the verification speed.
In DeTEC, blockchain serves as a distributed ledger to record the contributions of edge servers so as to reward them for their computations.

In \cite{xu2019become}, a blockchain-enabled computation offloading scheme named BeCome is proposed for IoT in MEC to obtain the optimal offloading scheme.
In this scheme, VM instances are used for provisioning physical resources in IoT, and a VM allocation ledger updating algorithm is presented to monitor resource utilization.
Considering the modelled total time consumption, total energy consumption and the level of load balance, a candidate computation offloading strategy generating algorithm is designed using NSGA-III.
Moreover, a simple additive weighting method and a multi-criteria decision making method are employed to calculate the highest utility value of the strategies; this means obtaining the optimal scheme among the candidates.
Blockchain is used in this scheme as a dynamic VM allocation ledger to record the resource utilization data of edge devices and the unoccupied VM instances in order to provide the integrity of data.
For edge computing in 5G environment, the same problem is also raised and solved with a blockchain-based solution in the same way \cite{xuxiaolong2019blockchain}.

\begin{table*}[]
	\caption{Computation Offloading for IBEC}
	\label{table9}
	\centering
	\renewcommand\arraystretch{1.5}
	\begin{tabular}{l<{\centering}|m{1.5cm}<{\centering}|m{4.5cm}|m{6.8cm}|m{1.5cm}<{\centering}}
		\hline
		Refs. & Applications      & \makecell[c]{Purposes}   & \makecell[c]{Contributions}  & Other Supporting Technologies  \\ \hline
		\cite{cui2019decentralized} & Computation Offloading  & Secure, efficient and rewarding tasks allocation system with low latency.  & Devising a task allocation platform called DeTEC with heuristic algorithm be the allocating solution and centralized / decentralized verification schemes.   & Heu-Alloc algorithm   \\ \cline{1-5}
		\cite{xu2019become} &Computation Offloading  & Reducing time and energy costs and achieving data integrity of task offloading process. & Proposing a blockchain-enabled computation offloading method named BeCome and employing NSGA-III to achieve optimal offloading strategy. & NSGA-III  \\ \cline{1-5}
		\cite{xuxiaolong2019blockchain} & Computation Offloading& Guaranteeing integral operating performance and data integrity for computation offloading. & Designing a computation offloading method including a blockchain-based edge computing framework and NSGA-III enabled optimal offloading strategy.  & NSGA-III, 5G  \\ \cline{1-5}
		\cite{zhang2019joint} & Computation Offloading & Optimizing the total cost of all mobile devices and ensuring each device to participate in computation offloading successfully if it needs. & Jointly studying the computation offloading and coin loaning problems in blockchain-enabled MEC. & No  \\ \cline{1-5}
		\cite{zhanglejun2021resource} & Computation Offloading & Improving system performance and QoE of users.  & Proposing a consortium blockchain-based edge computing system with a group-agent strategy, a task sorting algorithm and a content search scheme. & No  \\ \cline{1-5}
		\cite{chuchunghua2021task} & Computation Offloading & Improving mining task offloading performance. & Proposing a MEC-enabled mining task offloading scheme for a redesigned lightweight blockchain and designing a DL-based BDO solution. & DL \\ \cline{1-5}
		\cite{qiu2019online} &  Computation Offloading  & Improving comprehensive performance of computational offloading in mobile edge computing. & Proposing an adaptive genetic algorithm-enabled DRL solution for high performance of online computation offloading in blockchain-enabled MEC by taking both mining tasks and data processing tasks into consideration. & Genetic algorithm, DRL  \\ \hline
		
	\end{tabular}
\end{table*}

To minimize the total cost of the mobile equipment (ME) in edge computing, a blockchain-based joint computation offloading and coin loaning system is proposed in \cite{zhang2019joint}.
In this system, the problem that the ME either chooses to loan from banks and pay for task offloading to edge servers, or chooses to compute locally is formulated as a noncooperative game.
The developed smart contracts on computing resource trading and coin loaning are deployed and run in an on-chain manner.
Subsequently, the performance and QoE of users are improved by a series of optimizations as mentioned in \cite{zhanglejun2021resource}, which include a consortium blockchain-based edge computing system with a group-agent strategy for trust computing, a task sorting mechanism for resource allocation improvement, and a content search model based on popularity for search optimization.

In  \cite{chuchunghua2021task}, a MEC-enabled PoW task offloading scheme is proposed by redesigning a micro-blockchain for high offloading performance.
In this scheme, only the block headers are stored an on-chain way, and a lightweight account tree structure is illustrated and put into the block header to record and update account balances of mobile users.
Moreover, a DL-based block data offloading (BDO) algorithm is designed to achieve optimal offloading schedule in order to maximize the utility of users.
Furthermore,  the adopted neural network model is trained to optimize its weights and biases by minimizing a loss function.
In \cite{qiu2019online}, an online DRL-based computation offloading approach is designed to achieve policy self-adjustment and system long-term performance in a highly dynamic and complexity environment of blockchain-enabled MEC.
In this approach, the computation offloading problem is formulated as a MDP to obtain the optimal strategy for executing  the mining and computational tasks.
Moreover, the adaptive genetic algorithm (AGA) is applied in the proposed approach to avoid useless exploration and improve convergence speed.

The above-mentioned studies on computation offloading for IBEC are summarized in Table \ref{table9}.

\begin{table*}[]
	\caption{Security Strategies for IBEC}
	\label{table10}
	\centering
	\renewcommand\arraystretch{1.5}
	\begin{tabular}{l<{\centering}|m{1.7cm}<{\centering}|m{4.5cm}|m{6.8cm}|m{1.5cm}<{\centering}}
		\hline
		Refs. & Applications      & \makecell[c]{Purposes}   & \makecell[c]{Contributions}  & Other Supporting Technologies  \\ \hline
		
		\cite{kochovski2019trust} & Security and Privacy & Achieving trust of data, software and infrastructure management among multi-parties in decentralized scenarios .  &  Studying vital attributes of trust, introducing a blockchain-enabled trust management system and implementing PoC.     & No       \\ \cline{1-5}
		\cite{guo2019blockchain} & Security and Privacy & High-efficient and secure authentication of IoT data access and data sharing.  &  Proposing a decentralized and trusted authentication system based on blockchain and edge computing for IoT data storing and sharing. & No  \\ \cline{1-5}
		\cite{yang2020distributed} & Security and Privacy  & Achieving privacy preservation of network topology in heterogeneous MEC system.  &  Designing a blockchain-based MEC system which implements multiplex mutual trust networking and collaborative routing verification using the blooming filter. & 5G  \\ \cline{1-5}
		\cite{ernest2020privacy} & Security and Privacy  & Simultaneously achieving confidentiality and transparency in the IBEC.& Proposing an ECC-based privacy-enhancement scheme (PES). & No \\ \cline{1-5}
		\cite{qiao2019anonymous} & Security and Privacy & Secure data storing and data sharing. &  Proposing an anonymous data storage and transaction protocol, which generates ElGamal cryptosystem-based pseudo identities and blinding signature-based electronic cheques. & No  \\ \cline{1-5}
		\cite{jiang2019messaging} & Security and Privacy & Optimizing security and performance of M2M communication in IoT environment. & Establishing a hybrid messaging model where the IoT devices layer maintains multi-group private blockchains and edge servers layer maintains a public blockchain.  & No  \\ \cline{1-5}
		\cite{bai2019messaging} & Security and Privacy   & Designing a messaging protocol that fits the IBEC. & Comparing and analysing the MQTT and CoAP messaging protocols in IoT network and proposing a novel one by combining the two protocols. & No  \\ \cline{1-5}
		\cite{lu2020communication} & Security and Privacy  &  Achieving high-quality and high-reliability in MEC communication service.   &  Proposing a digital twin-enabled edge computing for network optimizing and asynchronous aggregation-based resource allocation scheme with blockchain for communication security and data privacy.  & FL, Digital Twin        \\ \cline{1-5}
		\cite{houlu2020design} & Security and Privacy  & Improving security of LoRa technology-based communication in IoT.  & Implementing a prototype LoRa system using edge computing and blockchain and analysing four security issues. & LoRa \\ \cline{1-5}
		\cite{mendki2020blockchain} & Security and Privacy & Addressing the security and privacy challenges of resource allocation. &  Reviewing challenges for computation resource lending, discussing the corresponding security and privacy technologies and proposing to use unidirectional payment channels for addressing the challenges.  & No   \\ \hline

	\end{tabular}
\end{table*}

\subsection{Security Strategies for IBEC}

Besides the characteristics of transparency, traceability and tamper-proofing of blockchain, the functions of access control and identity authentication can be implemented in form of smart contracts to achieve security and privacy in the IBEC system.

In \cite{kochovski2019trust}, a trust management architecture is proposed by using smart contracts and smart oracles to provide security services across the Edge-Fog-Cloud computing continuum to support development and operation of smart applications.
In this architecture, the smart oracles are used to assess and provide specific metrics for smart contracts,  which are implemented to monitor off-chain data and to select optimal fog nodes to reduce the cost and to improve QoS of the whole system.

In order to achieve secure authentication and collaborative sharing, a distributed and trusted authentication system is developed in \cite{guo2019blockchain} by combining blockchain and edge computing, in which an optimized PBFT consensus algorithm is proposed to help store authentication data and logs on blockchain.
Moreover, a domain name system (DNS)-based dynamic name resolution strategy is designed to provide authentication and data synchronization services for terminals, and an elliptic curve cryptography (ECC)-based cryptographic algorithm is designed to ensure anonymity and communication security between terminals and edge nodes. Furthermore, content caching strategy based on belief propagation (BP) algorithm is utilized to improve the downloading efficiency by cooperation.

To preserve network topology privacy, a consortium blockchain-based trusted MEC multi-domain collaboration architecture named BlockTC is proposed in \cite{yang2020distributed}, where SDN controllers of all domains are authenticated to own credible access identities, and are responsible to perform the routing verification consensus and maintain the blockchain.
Specifically, in order to balance radio resources, optical resources and edge servers' resources, a MEC collaboration routing algorithm is designed in BlockTC for SDN controllers to calculate the weights of links and servers and to choose the minimum routing link.

To protect the privacy of cross-domain routing in multi-domain MEC network, a consortium blockchain-based routing verification scheme, a network-driven collaboration routing verification (ND-CRV) scheme and a cloud-driven CRV (CD-CRV) scheme, are proposed in \cite{ernest2020privacy} by using the blooming filter to generate new requests for the subsequent SDN controllers without exposing topology privacy.
Moreover, to simultaneously achieve confidentiality and transparency, an ECC-based privacy-enhancement scheme (PES) in the IBEC environment is developed, which includes public key random generation and digital signature generation.

In \cite{qiao2019anonymous}, a data anonymous storage and transaction protocol is proposed by utilizing blockchain and edge computing to achieve the anonymity during processes of data storing and trading.
In the stage of data anonymous storage, the edge nodes generate pseudo identities for the collected data from IoT devices, encrypt the data using symmetric key, then store the data to the distributed hash table (DHT) and finally obtain the addresses.
In the stage of data anonymous transactions, the edge devices first register on blockchain and serve as data sellers;
when data buyer requests the transaction and sends a digital cheque, the seller delivers the encrypted data and
a bank transfers money with the cheque, where the information of requests and cheques are recorded on blockchain.

For the purpose of achieving secure and efficient messaging in the IBEC, a messaging model is developed in \cite{jiang2019messaging}.
The private blockchain and public blockchain are together utilized in the messaging model to respectively support the communications within an edge group and in the whole network.
Sequentially, a hybrid messaging method  is proposed in \cite{bai2019messaging} by integrating the Message Queuing Telemetry Transport (MQTT) and Constrained Application Protocol (CoAP), in which the less power-consuming CoAP protocol is used for resource-limited terminals to communicate with their edge servers, while the MQTT protocol is utilized for communication between edge / fog nodes.
To be closer to the practical edge networks with complexities and heterogeneity, the digital twin technology is introduced in edge networks \cite{lu2020communication}, where the permissioned blockchain-enabled FL scheme is proposed to enhance the communication security and data privacy.
In this method, the asynchronous model update and aggregation algorithm is adopted to reduce the delay of users, and a RL-based solution is designed to train the digital twin-enabled policy DNN to finally optimize user scheduling and bandwidth allocation.

As one of the low-power wide-area (LPWA) technologies, long-range (LoRa) has been widely applied in IoT environment to provide energy-efficient communications, although it is challenging to guarantee its security and efficiency.
In \cite{houlu2020design}, two blockchain ledgers are deployed in the proposed LoRa system, where LoRa gateways are responsible for constructing blocks using the Merkle tree and executing root value generation algorithms to maintain the network ledger. Another blockchain is regarded as application ledger that could only be accessed by authorized application servers. The security issues of application-layer, including denial-of-service (DoS) attack, SPOF, malicious LoRa gateway and malicious network server, are analyzed and demonstrated to be mitigated or eliminated in the proposed LoRa system.

Different from the above software-based security solutions, a hardware-based solution for the blockchain-enabled edge computing architecture is proposed in \cite{mendki2020blockchain} by using Intel Software Guard Extension (SGX) to realize data privacy, task and code tamper-proofing, incentive mechanism and execution in a sandboxed environment.

The above-mentioned studies on security strategies for IBEC are summarized in Table \ref{table10}.

\section{Challenges and Solutions}
This section presents the challenges and the potential solutions of the IBEC.

\subsection{Performance Scalability}

Performance scalability is one of the biggest challenges in blockchain application.
For instance, in Bitcoin ledger, blockchain is explicitly stipulated on its block size, consensus mechanism, broadcasting algorithm, \textsl{etc.} for its decentralization and security.
However, when the number of users surges, these regulations lead to a decrease in Bitcoin's throughput, \textsl{e.g.}, Bitcoin can only process 3 to 7 transactions per second simultaneously.

With respect to the IBEC, energy trading or knowledge trading are two major application scenarios in recent years.
One of the important functions of these two scenarios is to optimize the performance of public blockchains.
For instance, designing a novel lightweight blockchain or proposing a novel consensus mechanism to fulfill the specific demands.
However, these solutions do not meet the rapid growth of verification demands for blockchain transactions especially in IoT applications, which combine blockchain and edge computing.
Moreover, there is an urgent need of intelligent autonomy for the distributed networks of air-space-ground-sea integration areas in 6G and beyond context. These emerging domains require blockchain-based solutions and also pose higher demands for the performance of underlying blockchains. In other words, large volumes of data and huge numbers of devices in 6G context will pose a great challenge on existing blockchain systems in terms of performance scalability.
As a result, the performance scalability of the IBEC remains to be improved due to the performance scalability issue of blockchain.
Accordingly, how to improve the performance scalability of blockchain is a challenge, especially for the IBEC.

The solutions of blockchain expansion are fully summarized in \cite{sanka2021systematic, khan2021systematic, zhouqiheng2020solutions, kaur2020scalability, yangli2020overview}. These solutions mainly concentrate on on-chain expansions, off-chain expansions and the zero's layer ones.
On-chain expansions refer to the improvements of basic protocols on the data layer, consensus layer and network layer. The improvements of protocols including block size increment, the delay reduction of block broadcasting and gaining consensus, \textsl{etc.} Omnilegder \cite{kokoris2018omniledger}, Bitcoin-Cash \cite{kwon2019bitcoin}, Spectre \cite{sompolinsky2016spectre}, PBFT \cite{liwenyu2020scalablemulti} and DPoS \cite{zhengzibin2017overview} are the typical cases of on-chain expansions.

Off-chain expansions mainly lie in the improvements of the application layer. These countermeasures include state channel, side-chain, cross-chain and off-chain computation. Lightning Network \cite{poon2016bitcoin}, Raiden Network \cite{hees2016raiden}, Plasma \cite{poon2017plasma} and Cosmos \cite{kwon2019cosmos} are the typical cases of off-chain expansions.

The zero's layer expansions improve the performance scalability by optimizing the underlying data transfer protocols of blockchain, \textsl{e.g.}, Blockchain Distribution Network (BDN) utilized by bloXroute \cite{klarman2018bloxroute}.
However, due to the fact that the IBEC has the massively dispersed and small pieces of resources, it is challenging to fulfill the rising demands of IoT. In the future, being empowered by AI algorithms, lightweight and flexible blockchain expansion schemes for the resource-limited and heterogeneous edge devices should be designed to improve the performance scalability.

\subsection{Resource Management}

In the IBEC,  a large number of heterogeneous edge devices can collaborate with edge servers to serve nearby devices when they are idle.
Moreover, most edge devices are really resource-limited devices, which need to resort to the resources provided by edge servers or other devices to complete corresponding tasks. These cases will occur more commonly in 6G and beyond context, in which the global deep fusion of air, space, ground and sea networks will involve much more related networks, network carrying equipment and heterogeneous accessing devices. This will inevitably pose a new challenge for resource management for the IBEC.
Furthermore, many application scenarios of the IBEC focus on data sharing and knowledge discovery; this requires ML or other technologies to perform a lot of computation-intensive tasks.
In addition, maintaining the blockchain itself requires huge computational resources.
Therefore, resource management is a key requirement in the IBEC.

With the rapid development of  Internet of Everything \cite{2020Unmannedaerial}, more devices and facilities will be connected to the network.
The proliferation and heterogeneity of devices, and frequent online and offline of IoT devices increase the difficulty of resource integration and  management.
Therefore, how to optimize resource management and improve the QoS of the IBEC  is the key challenge.
An alternative solution is to design a more reasonable, practical and efficient bidding and pricing mechanism under a more realistic trading model, which can encourage devices to participate in the exchange of resources.
Moreover,  an AI-based edge intelligent resource management system with adaptive integration and allocation of resources should be proposed by utilizing emerging technologies (\textsl{e.g.}, FL) to cope with the highly dynamic and complexity of the IBEC.

\subsection{Security}

The IBEC brings security challenges, which exhibit in two aspects: (1) how to realize the security of smart contract; and (2) how to achieve efficient identification of edge devices.

The IBEC puts forth specific functions, such as resource allocation, identity authentication and access control, which are generally implemented in the form of smart contracts.
Despite the merits such as high efficiency, customization, flexibility, automation, \textsl{etc.}, smart contracts cause large economic accidents due to their security vulnerabilities, inalterable error codes, and malicious codes \cite{vacca2021systematic, tolmach2021survey,2019Anoverviewonsmart}.
Moreover, there is no effective legal supervision of smart contracts.
Therefore, it is a challenge to ensure that smart contracts are completely correct and effective, especially in the IBEC. One solution is to strengthen the security analysis of smart contracts by designing practical privacy preservation encrypted traffic inspection scheme on blockchains.
With this scheme, the abnormal and suspicious codes of smart contracts can be detected while preserving the privacy of smart contracts.

The identification of edge devices is another key issue to establish trust among devices or cross systems.
Since a large number of heterogeneous devices need to be authenticated in order to join various networks to participate in activities, thereby posing huge computing pressure on the edge side.
Moreover, there are diverse authentication standards.
Accordingly, it is still a challenge to design a lightweight, fast and consistent authentication mechanism.
The above issue may be potentially solved by designing a distributed IoT device identification mechanism, as well as
promoting the standardizations of IoT device identification protocol.

\subsection{Privacy Computing}

Data sharing and knowledge discovery are the basic requirements in many application scenarios (such as, IoV, IIoT, and smart healthcare) of the IBEC. For example, knowledge discovery requires a comprehensive analysis of a large number of patient records. Meanwhile, sharing the GPS positioning data corrects errors in assisted automatic driving.
Especially in future generation network environment, data generated by more equipment and devices will be more scattered, and the increasing number of stakeholders drives the great demand of data sharing.
Privacy preservation of the data in the IBEC system  is important, as a large amount of personally sensitive data is stored in the system.
However, existing works usually focus on the security of data, which is nevertheless limited by the underlying security mechanism of blockchains.
Moreover, traditional privacy preservation mechanisms are not suitable for edge devices due to the complexities.
Therefore, how to protect data privacy in the computing process is the key to widen the applications of the IBEC.

At this stage, technologies related to data privacy computing include privacy-enhanced computing (PEC), multi-party secure computing (MPSC), \textsl{etc}.
To apply these technologies efficiently in the IBEC,
it is necessary to consider issues of limited resources on edge devices and huge demands for computing offloading.
Therefore, lightweight, parallel and modular optimizations on PEC algorithms, MPSC algorithms or others that can protect data computing privacy may be new trends to improve the applicability and scalability of the IBEC system.

\section{Conclusion}
As two crucial technologies for IoT, both blockchain and edge computing have been used to realize secure and efficient resource management, computation offloading and data sharing.
This survey starts with a brief introduction of blockchain and edge computing and then presents the architecture of an IBEC system in IoT application scenarios.
We next classify and discuss the corresponding research issues and existing solutions, in perspectives of resource management, data management, computation offloading,  security and privacy, which have attracted most attentions in the areas of IoV, Smart Grid, smart healthcare, IIoT, \textsl{etc.}
Finally, we summarize research challenges, which have hindered the further large-scale deployments. We also discuss performance optimization for the IBEC, in terms of performance scalability, resource management, security, and privacy computing.

\bibliographystyle{IEEEtran}
\bibliography{reference}
\end{document}